 \documentclass[preprint,superscriptaddress,nofootinbib,aps]{revtex4}
 \usepackage{graphicx}
 \usepackage{graphics}
 \usepackage{bm}
 \begin{document}

 \title{Global analysis of $B$ ${\to}$ $PP$, $PV$ charmless
    decays with QCD factorization}
 \thanks{Talk given at the International Conference on Flavour
    Physics, Seoul, Korea, 6---11 October, 2003}
 \author{Dongsheng Du}
 \email[E-mail address: ]{duds@mail.ihep.ac.cn}
 \affiliation{Institute of High Energy Physics,
              Chinese Academy of Sciences,\\
              P.O.Box 918(4),
          Beijing 100039, P. R. China}
 \begin{abstract}
 The global analysis of $B$ ${\to}$ $PP$, $PV$ charmless decays
 with QCD factorization (QCDF) is presented. The predictions of
 QCDF are in good agreement with experiments. The best fitted CKM
 angel ${\gamma}$ is around $79^{\circ}$. The predicted branching
 ratios of $B$ ${\to}$ ${\pi}^{0}{\pi}^{0}$, ${\omega}K^{+}$,
 ${\omega}{\pi}^{+}$, ${\pi}^{+}K^{{\ast}0}$ etc. are also in good
 agreement with new data of {\sc BaBar} and {\sc Belle}.
 \end{abstract}

 \maketitle

 \section{Introduction}
 \label{sec1}
 The charmless two-body $B$ decays are very important for extracting CKM
 angles ${\alpha}$, ${\gamma}$ and for testing QCD. Up to now, {\sc BaBar}
 and {\sc Belle} have accumulated large set of data. It is highly
 interesting to analysis these data by using different theories and
 compare the data with theoretical predictions. QCD factorization
 \cite{du1} has been used to analysis $B$ ${\to}$ $PP$, $PV$ charmless
 decay data \cite{du2,du3} (here $P$ denotes pseudoscalar meson, $V$
 vector meson). With the data at that time, the theoretical results prefer
 a larger CKM angle ${\gamma}$. The QCD factorization (QCDF) predictions
 for some $B$ ${\to}$ $PV$ channels are only marginally consistent with
 the experimental data. Notice that the QCDF predictions contain large
 numerical uncertainties due to the CKM matrix elements, form factors, and
 annihilation parameters. Furthermore, the uncertainties of various decay
 channels are strongly correlated to each other. So we are stimulated to
 do a global analysis to check the consistency between the QCDF
 predictions and the experimental data. Beneke {\em et al.} \cite{du4}
 have done a gloabl analysis for $B$ ${\to}$ $PP$, including ${\pi}{\pi}$,
 ${\pi}K$ modes with QCDF approach. The results show a satisfactory
 agreement between QCDF predictions and data for  $B$ ${\to}$ $PP$
 branching fractions. But their fitted CKM angle ${\gamma}$ ${\sim}$
 $90^{\circ}$ which is not consistent with the standard CKM fit
 \cite{du5}: $37^{\circ}$ ${\le}$ ${\gamma}$ ${\le}$ $80^{\circ}$. Now
 {\sc BaBar} and {\sc Belle} have a lot of data on $B$ ${\to}$ $PV$. It is
 necessary to do the global fit to $B$ ${\to}$ $PP$ and $PV$ at the same
 time. We have done it and found that QCDF can fit all data on $B$ ${\to}$
 $PP$, $PV$. The fitted angle ${\gamma}$ ${\sim}$ $79^{\circ}$ which is
 consistent with the standard CKM global fit. We also have new predictions
 on $B$ ${\to}$ ${\pi}^{0}{\pi}^{0}$, $K^{+}K^{-}$,
 ${\pi}^{+}K^{{\ast}0}$, ${\omega}{\pi}^{+}$, ${\omega}K^{+}$.

 My talk is organized as follows: in Section \ref{sec2}, I give general
 remarks on QCD factorization. Section \ref{sec3} is devoted to the global
 analysis on $B$ ${\to}$ $PP$, $PV$ decays. In Section \ref{sec4}, I give
 the summary and conclusion.

 \section{General remarks on QCD factorization}
 \label{sec2}
 The low energy effective Hamiltonian for ${\vert}B{\vert}$ $=$ $1$ is
 \begin{eqnarray}
 {\cal H}_{eff} &=& \frac{G_{F}}{\sqrt{2}}
   {\sum\limits_{q=u,c}} v_{q} \Big\{
     C_{1}({\mu}) Q^{q}_{1}({\mu})
  +  C_{2}({\mu}) Q^{q}_{2}({\mu}) \nonumber \\
  &+& {\sum\limits_{k=3}^{10}} C_{k}({\mu}) Q_{k}({\mu})
  \Big\} + \text{H.c.} ,
 \label{eq1}
 \end{eqnarray}

 The decay amplitude for $B$ ${\to}$ $M_{1}M_{2}$ is
 \begin{eqnarray}
 {\cal A}(B{\to}M_{1}M_{2}) = \frac{G_{F}}{\sqrt{2}}
 \sum\limits_{i} \sum\limits_{q} C_{i}(\mu)
 {\langle}M_{1}M_{2}{\vert}Q_{i}(\mu){\vert}B{\rangle},
 \label{eq1-2}
 \end{eqnarray}

 The Wilson coefficients $C_{i}(\mu)$ include short distance effects above
 the scale ${\cal O}(m_{b})$ and are perturbatively calculable. The long
 distance effects (contributions from the scale below ${\cal O}(m_{b})$)
 are included in the hadronic matrix elements
 ${\langle}M_{1}M_{2}{\vert}Q_{i}(\mu){\vert}B{\rangle}$.

 For naive factorization (BSW),
 \begin{equation}
 \begin{array}{ccc}
 {\langle}M_{1}M_{2}{\vert}Q_{i}(\mu){\vert}B{\rangle}&{\sim}&
 {\langle}M_{1}{\vert}J_{1}{\vert}0{\rangle}
 {\langle}M_{2}{\vert}J_{2}{\vert}B{\rangle} \\
  & & \swarrow \ \ \ \ \ \ \ \ \ \searrow \\
  & \multicolumn{2}{c}{\rm decay\ constant\ \ \ \ \ form\ factor}
 \end{array}
 \label{eq2}
 \end{equation}
 There are no renormalization scheme and scale dependences. So the
 corresponding dependences of $C_{i}(\mu)$ connot be canceled and the
 ``non-factorizable'' contributions cannot be accounted for.

 In 1999, Beneke {\em et al.} \cite{du1} proposed a new factorization
 scheme based on QCD. That is so-called QCD factorization (QCDF). In this
 scheme, in the heavy quark limit,
 \begin{widetext}
 \begin{eqnarray}
     {\langle}M_{1}M_{2}{\vert}O_{i}{\vert}B{\rangle}
 &=& F^{B{\to}M_{2}}_{j}(q^{2}) {\int}_{0}^{1}dx T^{I}_{i}(x)
     {\Phi}_{M_{1}}(x) + ( M_{1} {\leftrightarrow} M_{2} ) \nonumber \\
 &+& \sum\limits_{j} {\int}_{0}^{1} d{\xi} {\int}_{0}^{1} dx
    {\int}_{0}^{1} dy T^{II}_{ij}({\xi},x,y) {\Phi}_{B}(\xi)
    {\Phi}_{M_{1}}(x) {\Phi}_{M_{2}}(y)
 + {\cal O}({\Lambda}_{QCD}/m_{b})
 \label{eq3}
 \end{eqnarray}
 \end{widetext}
 where $T^{I,II}_{i}$ denote hard scattering kernels which are calculable
 order by order in perturbation theory and include short distance
 effects. The long distance effects are included in decay constants, form
 factors and ling-cone distribution amplitudes. $M_{1}$ is a light meson
 or a charmonium state. $M_{2}$ (containts the spectator quark in $B$
 meson) is any light or heavy meson. If $M_{2}$ is heavy (for example, $D$
 meson), the second line in the Eq.(\ref{eq3}) is $1/m_{b}$ suppressed.

 \begin{figure*}
 \includegraphics{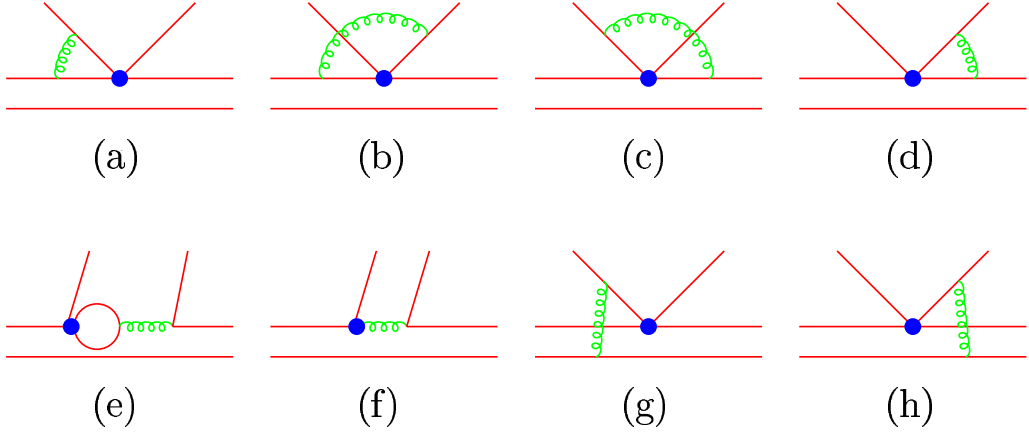}
 \caption{\label{fig1} $T^{I,II}$ at ${\cal O}({\alpha}_{s})$. The upward
 quark lines denote the ejected meson $M_{1}$}
 \end{figure*}

 \begin{figure*}[htb]
 \includegraphics{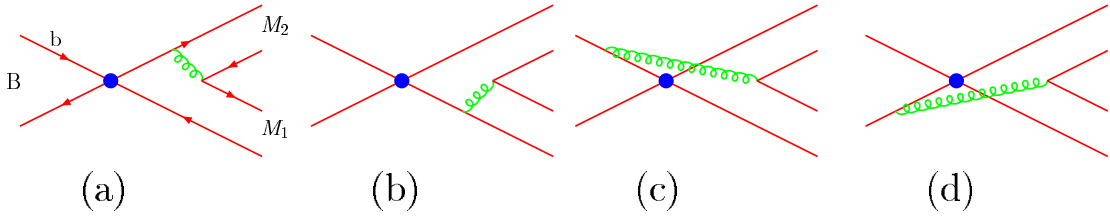}
 \caption{\label{fig2} annihilation contributions}
 \end{figure*}

 General observation:
 \begin{itemize}
 \item At the zeroth order of ${\alpha}_{s}$, it reduces to naive
       factorization
 \item At the higer order, the corrections can be computed systematically
 \item The renormalization scheme and scale dependence of
       ${\langle}Q_{i}{\rangle}$ is restored. In the heavy quark limit,
       the ``non-factorization'' contributions is calculable
       perturbatively. It does not need to introduce $N^{\rm eff}_{\rm c}$
 \item In heavy quark limit, strong phases arise solely from vertex and
       penguin corrections, so are at ${\cal O}({\alpha}_{s})$. That means
       that {\em strong phases are ${\alpha}_{s}$-suppressed}.
 \item Numerically, ${\Lambda}_{QCD}/m_{b}$ ${\sim}$ ${\cal O}({\alpha}_{s})$,
       so power corrections contribution to strong phases are also
       important. We do not have a systematic way to estimate power
       (${\Lambda}_{QCD}/m_{b}$) corrections. Thus {\em the calculation
       of strong phase is not reliable. So $CP$ asymmetry calculation is
       unreliable}.
 \item $W$-exchange \& $W$-annihilation diagrams are $1/m_{b}$ suppressed
 \item No long distance interactions between $M_{1}$ and ($BM_{2}$)
 \item $T^{I}$ includes: \\
       \begin{tabular}{ll}
       ${\alpha}_{s}^{0}:$ & tree diagram \\
       ${\alpha}_{s}:$     & non-factorizable gluon exchange \\
                           & vertex corrections (Fig.\ref{fig1}(a)-(d)) \\
                           & penguin corrections (Fig.\ref{fig1}(e)-(f)) \\
       ${\vdots}$          &
       \end{tabular}
 \item $T^{II}$ includes: \\
       \begin{tabular}{ll}
       ${\alpha}_{s}:$ & hard spectator scattering (Fig.\ref{fig1}(g)-(h)) \\
       ${\vdots}$      &
       \end{tabular}
 \end{itemize}
 Superficially, ${\Lambda}_{QCD}/m_{b}$ ${\sim}$ $1/15$ is a small number,
 But in some cases, such power suppression fails numerically. For example
 \begin{equation}
 {\langle}Q_{6}{\rangle}_{\rm F} = -2 \sum\limits_{q^{\prime}}
 {\langle}P_{1}{\vert}(\bar{q}q^{\prime})_{S-P}{\vert}0{\rangle}
 {\langle}P_{2}{\vert}({\bar{q}}^{\prime}b)_{S+P}{\vert}B{\rangle}
 \label{eq4}
 \end{equation}
 is always multiplied by a formally power suppressed but chirally enhanced
 factor $r_{\chi}$ $=$ $2{\mu}_{P}/m_{b}$ (${\mu}_{p}$ $=$
 $m_{P}^{2}/(m_{1}+m_{2})$, $m_{i}$ are current quark masses). In the
 heavy quark limit, $m_{b}$ ${\to}$ ${\infty}$, $r_{\chi}$ ${\to}$ $0$.
 But for $m_{b}$ ${\sim}$ 5GeV, $r_{\chi}$ ${\sim}$ ${\cal O}(1)$, no
 $1/m_{b}$ suppression. So we should consider the chirally enhanced power
 corrections. For example, in $B$ ${\to}$ ${\pi}K$, penguin is important.
 Dominant contribution to the amplitude ${\sim}$ $a_{4}+a_{6}r_{\chi}$,
 ($a_{4}$ ${\sim}$ $a_{6}r_{\chi}$). So chirally enhanced power
 correction term $a_{6}r_{\chi}$ cannot be neglected.

 Possible sources of power corrections:
 \begin{itemize}
 \item high twist wave functions;
 \item quark transverse momentum $k_{\bot}$;
 \item annihilation topology diagrams.
 \end{itemize}

 When we include chirally enhanced power corrections,
       the infrared (including soft and collinear) divergences for vertex
       corrections cancel only if ${\Phi}_{\sigma}(x)$ is symmetric, i.e.
       ${\Phi}_{\sigma}(x)$ $=$ ${\Phi}_{\sigma}(1-x)$ (${\Phi}_{\sigma}$,
       ${\Phi}_{p}$ are twist-3 wave functions).
       But even ${\Phi}_{\sigma}(x)$ $=$ ${\Phi}_{\sigma}(1-x)$, there is
       still logarithmic divergence for hard spectator scattering and
       annihilation topology. It violates factorization.
      \begin{eqnarray}
     && \!\!\!\! \!\!\!\! \!\!\!\! X_{H,A} = {\int}_{0}^{1}\frac{dx}{x} =
       {\ln}\frac{m_{B}}{{\Lambda}_{h}} (1+{\rho}_{H,A}e^{i{\phi}_{H,A}})
        \nonumber \\
     && \!\!\!\! \!\!\!\! \!\!\!\! {\Lambda}_{h}{\sim} 0.5{\rm GeV}, \ \
        0<{\rho}_{H,A}<1, \ \ 0^{\circ}{\le}{\phi}_{H,A}<360^{\circ}
     \end{eqnarray}
       Other power corrections maybe unimportant based on Renormalon
       Calculus estimation \cite{du7}.

 \section{Global analysis of $B$ ${\to}$ $PP$, $PV$}
 \label{sec3}
 The decay amplitudes of $B$ ${\to}$ $PP$, $PV$ can be written as
 \begin{widetext}
 \begin{eqnarray}
    {\cal A}(B{\to}M_{1}M_{2})
 =  \frac{G_{F}}{\sqrt{2}} \sum\limits_{i} v_{q} a_{i}^{p}
    {\langle}M_{1}{\vert}J_{1}{\vert}0{\rangle}
    {\langle}M_{2}{\vert}J_{2}{\vert}B{\rangle}
 + \frac{G_{F}}{\sqrt{2}} f_{B} f_{M_{1}} f_{M_{2}}
     \sum\limits_{i} v_{q} b_{i}
 \end{eqnarray}
 where $b_{i}$ $=$ $b_{i}(X_{A},{\alpha}_{s})$ $=$ ${\alpha}_{s}f(X_{A})$
 is related to the contributions of annihilation topology.

 We use {\sc CKMFitter} package \cite{du8} developed for global analysis
 of $B$ ${\to}$ $PP$ and enlarged it to include $B$ ${\to}$ $PV$.
 The {\sc RFit} scheme is implemented for statistical treatment, assuming
 the experimental errors to be purely Gaussian, while {\em quantites that
 cannot be computed precisely are bound to remain within their predefined
 allowed ranges}.

 The input parameters are as follows:
 \begin{enumerate}
 \item two-parton light-cone distribution amplitudes. \\
       \begin{itemize}
       \item for pseudoscalar meson
       \begin{eqnarray}& &
     {\langle}P(k){\vert}{\bar{q}}_{\alpha}(z_{2})q_{\beta}(z_{1})
     {\vert}0{\rangle} \nonumber \\
     &=& \frac{if_{P}}{4}{\int}_{0}^{1}dx
     e^{i(xk{\cdot}z_{2}+\bar{x}k{\cdot}z_{1})}
     \Big\{ k\!\!\!\slash {\gamma}_{5} {\Phi}_{P}(x)
      \ - {\mu}_{P} {\gamma}_{5} \Big[ {\Phi}_{P}^{p}(x)
        - {\sigma}_{{\mu}{\nu}}k^{\mu}z^{\nu}
      \frac{{\Phi}_{P}^{\sigma}(x)}{6}
           \Big] \Big\}_{{\beta}{\alpha}}
       \label{eq7}
       \end{eqnarray}
       where $z$ $=$ $z_{1}-z_{2}$.
       \item for longitudinally polarized vector meson
       \begin{eqnarray}
       {\langle}V(k,{\lambda}){\vert}{\bar{q}}_{\alpha}(z_{2})
          q_{\beta}(z_{1}){\vert}0{\rangle}
       = -\frac{if_{V}}{4} m_{V} {\int}_{0}^{1}\!\! dx\,
     e^{i(xk{\cdot}z_{2}+\bar{x}k{\cdot}z_{1})} \Big\{ k_{\mu}
    \frac{{\varepsilon}_{\lambda}^{\ast}{\cdot}z}{k{\cdot}z}
       {\Phi}^{V}_{\|}(x) \big\}_{{\beta}{\alpha}},
       \label{eq8}
       \end{eqnarray}
       for vector meson, the contributions of the twist-3 LCDAs are doubly
       suppressed, so can be safely disregarded.
       \item In order to reduce the number of input parameters, we use the
       asymptotic light-cone distribution amplitudes for final light mesons
       instead of Gegenbauer polynomials.
       \begin{equation}
       {\Phi}_{P}(x)=6x\bar{x}, \ \ \
       {\Phi}_{P}^{\sigma}(x)=6x\bar{x}, \ \ \
       {\Phi}_{P}^{p}(x)=1, \ \ \
       {\Phi}^{V}_{\|}(x)=6x\bar{x},
       \label{eq9}
       \end{equation}
      \item for $B$ meson, its wave function appears only in the
      contributions of hard spectator scattering \cite{du4}
      \begin{eqnarray}
      {\int}_{0}^{1}\frac{d{\xi}}{\xi}{\Phi}_{B}(\xi)
         {\equiv} \frac{m_{B}}{{\lambda}_{B}},\ \ \ \ \
      {\lambda}_{B}=(350{\pm}150)\,{\rm MeV},
       \end{eqnarray}
       \end{itemize}
 \item chirally enhanced factor \& annihilation-related parameters
      \begin{eqnarray}
      &&r_{\chi}^{\pi} {\simeq} r_{\chi}^{\eta} {\simeq} r_{\chi}^{K}\
        =\ \frac{2m_{K}^{2}}{m_{s}m_{b}}, \ \ \
        m_{b}(m_{b})=4.2\,{\rm GeV}, \ \ \
        m_{s}(2\,{\rm GeV})=(110{\pm}25)\,{\rm MeV}, \nonumber \\
      &&X_{H,A}^{PP}={\ln}\frac{m_{B}}{{\Lambda}_{h}}
           (1+{\rho}_{H,A}^{PP}e^{i{\phi}_{H,A}^{PP}}), \ \ \ \ \ \ \ \ \
        X_{H,A}^{PV}={\ln}\frac{m_{B}}{{\Lambda}_{h}}
           (1+{\rho}_{H,A}^{PV}e^{i{\phi}_{H,A}^{PV}})
      \label{eq10}
      \end{eqnarray}
     (note: $X_{H}$, $X_{A}$ for $B$ ${\to}$ $PP$ are different
            for $B$ ${\to}$ $PV$)
 \item Decay constants and form factors
      \begin{equation}
      \begin{array}{l}
      \begin{array}{llll}
      f_{\pi}=131\,{\rm MeV}, &
      f_{K}=160\,{\rm MeV}, &
      f_{K^{\ast}}=214\,{\rm MeV}, &
      f_{\rho}=210\,{\rm MeV}, \\
      f_{\omega}=195\,{\rm MeV}, &
      f_{\phi}=233\,{\rm MeV}, &
      f_{B}=180\,{\rm MeV}, &
      {\phi}=39.3^{\circ}, \\
      f_{q}=1.07f_{\pi}, &
      f_{s}=(1.34{\pm}0.06)f_{\pi}, &
      f_{\eta}^{q}=f_{q}{\cos}{\phi}, &
      f_{\eta}^{q}=-f_{s}{\sin}{\phi},
      \end{array} \\
      \begin{array}{lll}
      F_{0,1}^{B{\to}{\pi}}(0)=0.28{\pm}0.05, &
      A_{0}^{B{\to}{\rho}}(0)=0.30{\pm}0.05, &
      R_{{\pi}K}{\equiv}\frac{f_{\pi}F^{B{\to}K}}{f_{K}F^{B{\to}{\pi}}}
         =0.9{\pm}0.1, \\
      A_{0}^{B{\to}{\omega}}(0)=A_{0}^{B{\to}{\rho}}(0), & &
      F_{0,1}^{B{\to}{\eta}}=F_{0,1}^{B{\to}{\pi}}
          \left(\frac{{\cos}{\theta}_{8}}{\sqrt{6}}
               -\frac{{\sin}{\theta}_{0}}{\sqrt{3}}\right),
      \end{array} \\
      \begin{array}{lll}
      f_{q}{\to}{\vert}{\eta}_{q}{\rangle} =\frac{{\vert}u\bar{u}{\rangle}
                +{\vert}d\bar{d}{\rangle}}{\sqrt{2}}, &
      {\langle}0{\vert}s{\gamma}_{5}\bar{s}{\vert}{\eta}{\rangle}
    =-i\frac{m_{\eta}^{2}}{2m_{s}}\left(f_{\eta}^{s}-f_{\eta}^{u}\right) &
      {\theta}_{8}={\phi}-{\arctan}(\sqrt{2}f_{q}/f_{s}), \\
      f_{s}{\to}{\vert}{\eta}_{s}{\rangle}={\vert}s\bar{s}{\rangle}, &
      \frac{{\langle}0{\vert}u{\gamma}_{5}\bar{u}{\vert}{\eta}{\rangle}}
           {{\langle}0{\vert}s{\gamma}_{5}\bar{s}{\vert}{\eta}{\rangle}}
     =\frac{f_{\eta}^{u}}{f_{\eta}^{s}}, &
      {\theta}_{0}={\phi}-{\arctan}(\sqrt{2}f_{s}/f_{q}),
      \end{array}
      \end{array}
      \label{eq11}
      \end{equation}
 \item CKM parameters\\
       We use Wolfenstein parameterization for the CKM matrix and take
      \begin{equation}
       A=0.835, \ \ \ \ \
       {\lambda}=0.22, \ \ \ \ \
       {\vert}V_{ub}{\vert}=3.49{\pm}0.24_{\rm exp}{\pm}0.55_{\rm theo},
      \end{equation}
      and use the measured branching fractions listed in Table.\ref{tab1}
      as input. We did not use the branching ratios in Table.\ref{tab2}
      as input. The reason is that the errors of the data are large and
      the glue content of ${\eta}^{(\prime)}$-meson cannot be treated
      neatly. Detailed discussion can be found in our published paper
      \cite{du9}. For experimental constraints of $CP$ asymmetries, the
      corresponding QCDF predictions are not reliable. Thus we do not use
      measured $CP$ asymmetries as input in our global fit.
 \end{enumerate}
 \begin{table*}[htb]
 \begin{center}
 \caption{Experimental data used in the global fit}
 \label{tab1}
 \begin{tabular}{l|c|c|c|c} \hline
 \multicolumn{1}{c|}{${\cal B}r{\times}10^{6}$}
 & CLEO & {\sc BaBar} & {\sc Belle}
 & \begin{tabular}{c} weighted \\ average \end{tabular} \\ \hline
   $B^{0}$ ${\to}$ ${\pi}^{+}{\pi}^{-}$
 & $4.3^{+1.6}_{-1.4}{\pm}0.5$
 & $4.7{\pm}0.6{\pm}0.2$
 & $5.4{\pm}1.2{\pm}0.5$
 & $4.77{\pm}0.54$ \\ \hline
   $B^{+}$ ${\to}$ ${\pi}^{+}{\pi}^{0}$
 & $5.4^{+2.1}_{-2.0}{\pm}1.5$
 & $5.5^{+1.0}_{-0.9}{\pm}0.6$
 & $7.4^{+2.3}_{-2.2}{\pm}0.9$
 & $5.78{\pm}0.95$ \\ \hline
   $B^{0}$ ${\to}$ $K^{+}{\pi}^{-}$
 & $17.2^{+2.5}_{-2.4}{\pm}1.2$
 & $17.9{\pm}0.9{\pm}0.7$
 & $22.5{\pm}1.9{\pm}1.8$
 & $18.5{\pm}1.0$ \\ \hline
   $B^{+}$ ${\to}$ $K^{+}{\pi}^{0}$
 & $11.6^{+3.0+1.4}_{-2.7-1.3}$
 & $12.8^{+1.2}_{-1.1}{\pm}1.0$
 & $13.0^{+2.5}_{-2.4}{\pm}1.3$
 & $12.7{\pm}1.2$ \\ \hline
   $B^{+}$ ${\to}$ $K^{0}{\pi}^{+}$
 & $18.2^{+4.6}_{-4.0}{\pm}1.6$
 & $17.5^{+1.8}_{-1.7}{\pm}1.3$
 & $19.4^{+3.1}_{-3.0}{\pm}1.6$
 & $18.1{\pm}1.7$ \\ \hline
   $B^{0}$ ${\to}$ $K^{0}{\pi}^{0}$
 & $14.6^{+5.9+2.4}_{-5.1-3.3}$
 & $10.4{\pm}1.5{\pm}0.8$
 & $8.0^{+3.3}_{-3.1}{\pm}1.6$
 & $10.2{\pm}1.5$ \\ \hline
   $B^{+}$ ${\to}$ ${\eta}{\pi}^{+}$
 & $<5.7$
 & $<5.2$
 & $5.3^{+2.0}_{-1.7}\,(<8.2)$
 & $<5.2$ \\ \hline
   $B^{0}$ ${\to}$ ${\pi}^{\pm}{\rho}^{\mp}$
 & $27.6^{+8.4}_{-7.4}{\pm}4.2$
 & $28.9{\pm}5.4{\pm}4.3$
 & $20.8^{+6.0+2.8}_{-6.3-3.1}$
 & $25.4{\pm}4.3$ \\ \hline
   $B^{+}$ ${\to}$ ${\pi}^{+}{\rho}^{0}$
 & $10.4^{+3.3}_{-3.4}{\pm}2.1$
 & $<39$
 & $8.0^{+2.3}_{-2.0}{\pm}0.7$
 & $8.6{\pm}2.0$ \\ \hline
   $B^{0}$ ${\to}$ $K^{+}{\rho}^{-}$
 & $16.0^{+7.6}_{-6.4}{\pm}2.8$ &
 & $11.2^{+5.9+1.9}_{-5.6-1.8}$
 & $13.1{\pm}4.7$ \\ \hline
   $B^{+}$ ${\to}$ ${\phi}K^{+}$
 & $5.5^{+2.1}_{-1.8}{\pm}0.6$
 & $9.2{\pm}1.0{\pm}0.8$
 & $10.7{\pm}1.0^{+0.9}_{-1.6}$
 & $8.9{\pm}1.0$ \\ \hline
   $B^{0}$ ${\to}$ ${\phi}K^{0}$
 & $5.4^{+3.7}_{-2.7}{\pm}0.7$
 & $8.7^{+1.7}_{-1.5}{\pm}0.9$
 & $10.0^{+1.9+0.9}_{-1.7-1.3}$
 & $8.6{\pm}1.3$ \\ \hline
   $B^{+}$ ${\to}$ ${\eta}{\rho}^{+}$
 & $<10$ &
 & $<6.2$
 & $<6.2$ \\ \hline
   $B^{0}$ ${\to}$ ${\omega}K^{0}$
 & $<21$
 & $5.9^{+1.7}_{-1.5}{\pm}0.9$ &
 & $5.9{\pm}1.9$ \\ \hline
 \end{tabular}

 \caption{data not used in the global fit}
 \label{tab2}
 \begin{tabular}{l|c|c|c|c} \hline
 \multicolumn{1}{c|}{${\cal B}r{\times}10^{6}$}
 & CLEO & {\sc BaBar} & {\sc Belle}
 & \begin{tabular}{c} weighted \\ average \end{tabular} \\ \hline
   $B^{+}$ ${\to}$ ${\eta}K^{+}$
 & $<6.9$ & $<6.4$ & $5.2^{+1.7}_{-1.5}\,(<7.7)$
 & $<6.4$ \\ \hline
   $B^{+}$ ${\to}$ ${\pi}^{+}K^{{\ast}0}$
 & $<16$
 & $15.5{\pm}3.4{\pm}1.8$
 & $16.2^{+4.1}_{-3.8}{\pm}2.4$
 & $15.8{\pm}3.0$ \\ \hline
   $B^{0}$ ${\to}$ ${\pi}^{-}K^{{\ast}+}$
 & $16^{+6}_{-5}{\pm}2$ &
 & $26.0{\pm}8.3{\pm}3.5$
 & $19.0{\pm}4.9$ \\ \hline
   $B^{+}$ ${\to}$ ${\eta}K^{{\ast}+}$
 & $26.4^{+9.6}_{-8.2}{\pm}3.3$
 & $22.1^{+11.1}_{-9.2}{\pm}3.3$
 & $26.5^{+7.8}_{-7.0}{\pm}3.0$
 & $25.4{\pm}5.3$ \\ \hline
   $B^{0}$ ${\to}$ ${\eta}K^{{\ast}0}$
 & $13.8^{+5.5}_{-4.6}{\pm}1.6$
 & $19.8^{+6.5}_{-5.6}{\pm}1.7$
 & $16.5^{+4.6}_{-4.2}{\pm}1.2$
 & $16.4{\pm}3.0$ \\ \hline
   $B^{+}$ ${\to}$ ${\omega}K^{+}$
 & $<8$ & $<4$ & $9.2^{+2.6}_{-2.3}{\pm}1.0$
 & \\ \hline
   $B^{+}$ ${\to}$ ${\omega}{\pi}^{+}$
 & $11.3^{+3.3}_{-2.9}{\pm}1.5$
 & $6.6^{+2.1}_{-1.8}{\pm}0.7$
 & $<8.2$ & \\ \hline
 \end{tabular}
 \end{center}
 \end{table*}
 \end{widetext}

 \begin{table}[htb]
 \begin{center}
 \caption{Fit1 and Fit2 mean the best fit value with and without the
 contribution of the chirally enhanced hard spectator and annihilation
 topology, respectively.}
 \label{tab3}
  \begin{tabular}{l|c|c|c} \hline
  \multicolumn{1}{c|}{modes} & Exp. & Fit1 & Fit2 \\ \hline
   $B^{0}$ ${\to}$ ${\pi}^{+}{\pi}^{-}$
 & $4.77{\pm}0.54$ & 4.82 & 5.68 \\
   $B^{+}$ ${\to}$ ${\pi}^{+}{\pi}^{0}$
 & $5.78{\pm}0.95$ & 5.35 & 3.25 \\
   $B^{0}$ ${\to}$ $K^{+}{\pi}^{-}$
 & $18.5{\pm}1.0$  & 19.0 & 18.8 \\
   $B^{+}$ ${\to}$ $K^{+}{\pi}^{0}$
 & $12.7{\pm}1.2$  & 11.4 & 12.6 \\
   $B^{+}$ ${\to}$ $K^{0}{\pi}^{+}$
 & $18.1{\pm}1.7$  & 20.1 & 20.2 \\
   $B^{0}$ ${\to}$ $K^{0}{\pi}^{0}$
 & $10.2{\pm}1.5$  & 8.2 & 7.3 \\
   $B^{+}$ ${\to}$ ${\eta}{\pi}^{+}$
 & $<5.2$          & 2.8 & 1.8 \\
   $B^{0}$ ${\to}$ ${\pi}^{\pm}{\rho}^{\mp}$
 & $25.4{\pm}4.3$  & 26.7 & 29.5 \\
   $B^{+}$ ${\to}$ ${\pi}^{+}{\rho}^{0}$
 & $8.6{\pm}2.0$   & 8.9 & 8.5 \\
   $B^{0}$ ${\to}$ $K^{+}{\rho}^{-}$
 & $13.1{\pm}4.7$  & 12.1 & 5.1 \\
   $B^{+}$ ${\to}$ ${\phi}K^{+}$
 & $8.9{\pm}1.0$   & 8.9 & 7.1 \\
   $B^{0}$ ${\to}$ ${\phi}K^{0}$
 & $8.6{\pm}1.3$   & 8.4 & 6.7 \\
   $B^{+}$ ${\to}$ ${\eta}{\rho}^{+}$
 & $<6.2$          & 4.6 & 3.8 \\
   $B^{0}$ ${\to}$ ${\omega}K^{0}$
 & $5.9{\pm}1.9$   & 6.3 & 1.2 \\ \hline
 \end{tabular}
 \end{center}
 \end{table}

 \begin{figure*}
 \includegraphics{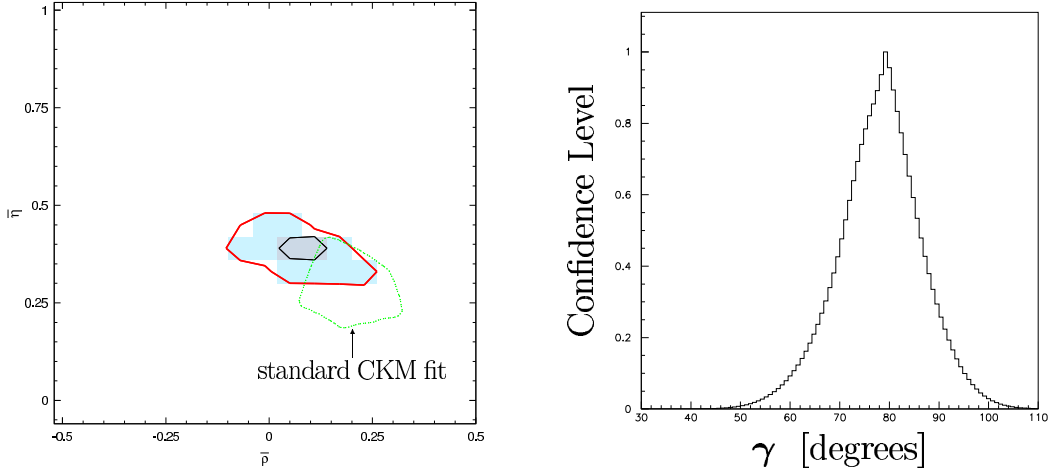}
 \caption{\label{fig3} left: the contours indicate ${\ge}90\%$ C.L.\&
  ${\ge}5\%$ C.L. right: C.L. of angle ${\gamma}$.}
 \end{figure*}

 \begin{figure*}
 \includegraphics{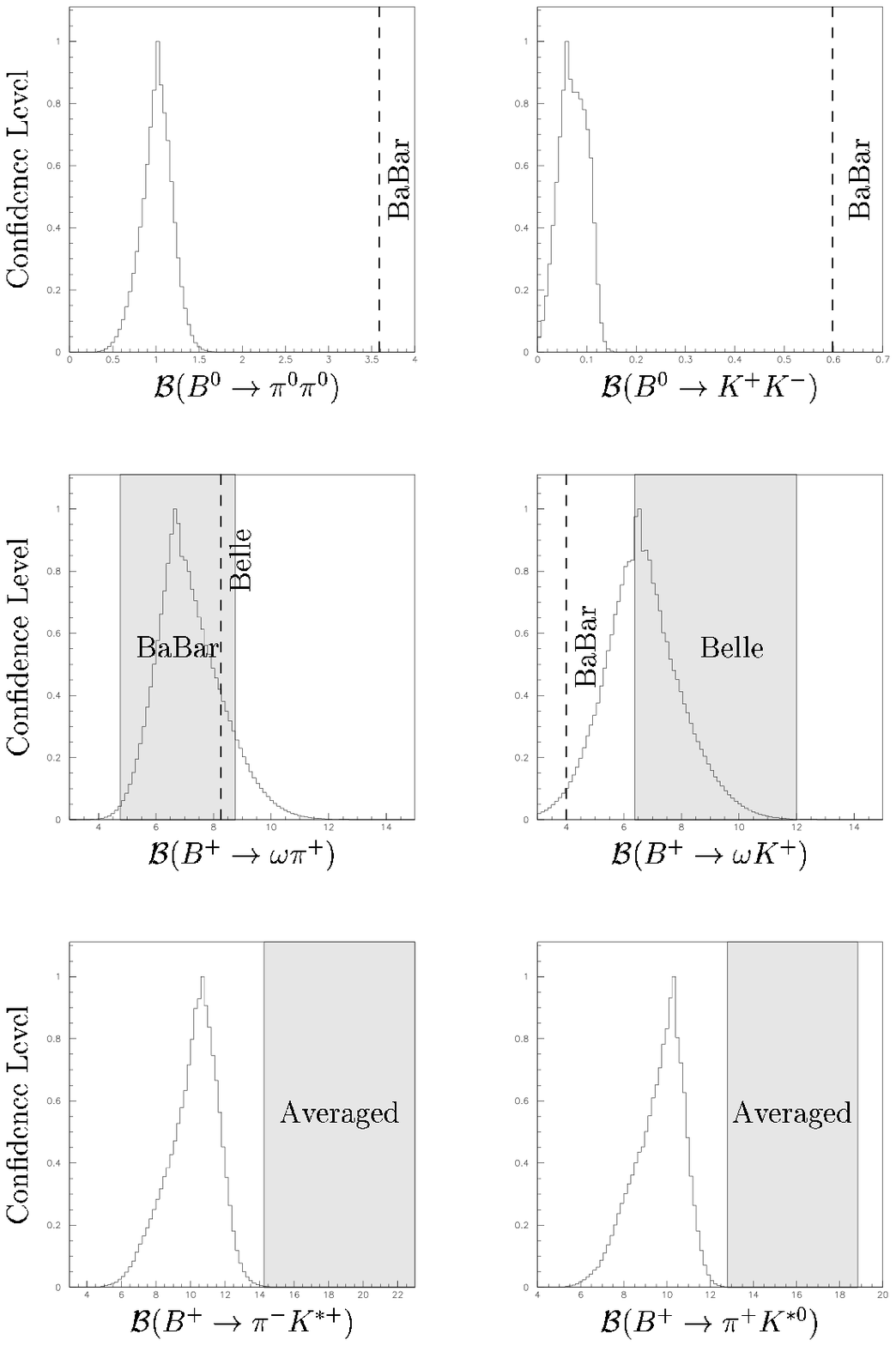}
 \caption{\label{fig4} The confidence levels of the fitted branching
  fractions (in the unit of $10^{-6}$).}
 \end{figure*}

 For $B$ ${\to}$ ${\pi}{\pi}$, ${\pi}K$, the fit is sensitive to
 ${\vert}V_{ub}{\vert}$, ${\gamma}\,({\rho},{\eta})$, $F^{B{\to}{\pi}}$,
 $F^{B{\to}K}$, $X_{A}$, $f_{B}/{\lambda}_{B}$, $m_{s}$. Including seven
 $B {\to} PV$ decay channels, only $A_{0}^{B{\to}{\rho}}$,
 $X_{A}^{PV}$ are newly involved sensitive parameters. So including $B$
 ${\to}$ $PV$ we shall have more stringent test of QCDF. Our
 best fit of ${\gamma}$ ${\sim}$ $79^{\circ}$ (see Figure.\ref{fig3})
 which is consistent with recent fit results $37^{\circ}$ $<$ ${\gamma}$
 $<$ $80^{\circ}$ \cite{du8}. The best fitted results are listed in
 Table.\ref{tab3}.

 The confidence levels of the fitted branching ratios are presented in
 Figure.\ref{fig4}. From Figure.\ref{fig4} we can see that
 \begin{itemize}
 \item For $B$ ${\to}$ ${\pi}^{0}{\pi}^{0}$, the best fit is around
       $1{\times}10^{-6}$ while the {\sc BaBar} and {\sc Belle} averaged
       measurement is $(1.90{\pm}0.49){\times}10^{-6}$.
 \item For $B^{+}$ ${\to}$ ${\omega}K^{+}$, the best fit is
       $6.25{\times}10^{-6}$ while the {\sc Belle} measurement is
       $(9.2^{+2.6}_{-2.3}{\pm}1.0){\times}10^{-6}$, and the CLEO
       measurement is $<8{\times}10^{-6}$.
 \item For $B^{+}$ ${\to}$ ${\omega}{\pi}^{+}$, the best fit is
       $6.66{\times}10^{-6}$ while the {\sc BaBar} measurement is
       $(6.6^{+2.1}_{-1.8}{\pm}0.7){\times}10^{-6}$, and the {\sc Belle}
       measurement is $<8.2{\times}10^{-6}$.
 \item For $B^{+}$ ${\to}$ ${\pi}^{+}K^{{\ast}0}$, the best fit ${\sim}$
       $10{\times}10^{-6}$ while the {\sc BaBar} preliminary measurement
       reported at {\sc LP03} is
       $(10.3{\pm}1.2^{+1.0}_{-2.7}){\times}10^{-6}$.
 \end{itemize}
 On the whole, our global fit is successful.

 \section{Summary \& Conclusion}
 \label{sec4}
 \begin{itemize}
  \item QCD factorization is a promising method for charmless two-body
        $B$ decays, which are cruical for the determination of the
        unitarity triangle.
  \item We enlarged the {\sc CKMFitter} package to include $B$ ${\to}$
        $PV$ charmless decay channels and did a global analysis. It is
        shown that the QCDF predictions are basically in good agreement
        with the experiments.
  \item We obtain ${\gamma}$ ${\sim}$ $79^{\circ}$, consistent with the
        CKM global fit.
  \item For $B$ ${\to}$ ${\pi}^{0}{\pi}^{0}$, the best fit is around
       $1{\times}10^{-6}$ while the {\sc BaBar} and {\sc Belle} averaged
       measurement is $(1.90{\pm}0.49){\times}10^{-6}$.
 \item For $B^{+}$ ${\to}$ ${\omega}K^{+}$, the best fit is
       $6.25{\times}10^{-6}$ while the {\sc Belle} measurement is
       $(9.2^{+2.6}_{-2.3}{\pm}1.0){\times}10^{-6}$, and the CLEO
       measurement is $<8{\times}10^{-6}$.
 \item For $B^{+}$ ${\to}$ ${\omega}{\pi}^{+}$, the best fit is
       $6.66{\times}10^{-6}$ while the {\sc BaBar} measurement is
       $(6.6^{+2.1}_{-1.8}{\pm}0.7){\times}10^{-6}$, and the {\sc Belle}
       measurement is $<8.2{\times}10^{-6}$.
 \item For $B^{+}$ ${\to}$ ${\pi}^{+}K^{{\ast}0}$, the best fit ${\sim}$
       $10{\times}10^{-6}$ while the {\sc BaBar} preliminary measurement
       reported at {\sc LP03} is
       $(10.3{\pm}1.2^{+1.0}_{-2.7}){\times}10^{-6}$.
 \end{itemize}
 The results are already published in \cite{du9}. I thank
 Drs. Junfeng Sun, Deshan Yang and Guohuai Zhu for their collaboration.


 \end{document}